\documentclass[intlimits,twoside,a4paper]{article}
\usepackage{amsmath,amssymb}
\usepackage{graphicx}
\usepackage{wrapfig}
\usepackage[T2A]{fontenc}
\usepackage[cp1251]{inputenc}

\usepackage[eqsecnum]{cmpj2}

\issue{2015}{18}{1}{13601}
\doinumber{10.5488/CMP.18.13601}

\title[Energetics of ion competition in Na channel]%
{Energetics of ion competition in the DEKA selectivity filter of neuronal sodium channels%
}
\author[D. Boda \textsl{et al.}]{D. Boda\refaddr{label4}\footnote{Author for correspondence: boda@almos.vein.hu}\,, G. Leaf\refaddr{label1},
J. Fonseca\refaddr{label2}, B. Eisenberg\refaddr{label3}}
\addresses{
\addr{label4} Department of Physical Chemistry, University of Pannonia, P. O. Box 158, Veszpr\'em, H-8201, Hungary
\addr{label1} Mathematics and Computer Science Division, Argonne National Laboratory, 9700 South Cass Avenue, \\ Argonne, IL 60439, USA
 \addr{label2} The Network for Computational Nanotechnology, Purdue University, 1205 West State Street, \\ West Lafayette, IN 47907, USA
 \addr{label3} Department of Molecular Biophysics and Physiology, Rush University Medical Center, 1750 West Harrison St., \\ Chicago, IL 60612 USA
}

\authorcopyright{D. Boda, G. Leaf, J. Fonseca, B. Eisenberg, 2015}
\date{Received September 2, 2014}

\begin{document}

\maketitle

\begin{abstract}
The energetics of ionic selectivity in the neuronal sodium channels is studied.
A simple model constructed for the selectivity filter of the channel is used.
The selectivity filter of this channel type contains aspartate (D), glutamate (E), lysine (K), and alanine (A) residues (the DEKA locus).
We use Grand Canonical Monte Carlo simulations to compute equilibrium binding selectivity in the selectivity filter and to obtain various terms of the excess chemical potential from a particle insertion procedure based on Widom's method.
We show that K$^{+}$ ions in competition with Na$^{+}$ are efficiently excluded from the selectivity filter due to entropic hard sphere exclusion.
The dielectric constant of protein has no effect on this selectivity.
Ca$^{2+}$ ions, on the other hand, are excluded from the filter due to a free energetic penalty which is enhanced by the low dielectric constant of protein.

\pacs 61.20.Qg, 68.03.-g, 81.05.Rm, 61.20.Ja, 07.05.Tp
\end{abstract}

\section{Introduction}
\label{sec:intro}

Sodium (Na) channels can be categorized on the basis of their function, the cell in which they are found, structure of the protein (both secondary and tertiary), and the structure of the selectivity filter (SF).
The SF is a narrow region of the permeation pathway, where the channel discriminates between different ions.
The selectivity properties of different channels primarily depend on the type of amino acid motifs present in their SF.

The two most widely studied classes of Na channels are the neuronal (this is the one studied here) and bacterial Na channels.
The SF of neuronal Na channels has a DEKA locus made of aspartate (D), glutamate (E), lysine (K), and alanine (A) residues.
On the basis of their homology with L-type calcium (Ca) channels \cite{2000_koch_jbc_34493}, these amino acids seem to face the permeation pathway.
The accurate structure of the DEKA Na channels is still unknown, so theoretical studies are restricted to using models based on homologies of the known structures or on reduced models based on minimal structural information available.
This is the approach used in this work, while the minimal structural information is that the SF has the DEKA locus.

The bacterial Na channels, on the other hand, have X-ray structures measured recently \cite{2012_mccusker_nc_1102,2011_payandeh_n_353,2013_stock_jpcb_3782,2013_ulmschneider_pnasusa_6364}.
These channels include NavMs \cite{2012_mccusker_nc_1102,2013_ulmschneider_pnasusa_6364}, NavAb \cite{2011_payandeh_n_353,2013_stock_jpcb_3782}, and  NaChBac \cite{2001_ren_s_2372}.
The structure of a Ca$^{2+}$-selective mutant of NavAb is also available \cite{2014_tang_n_56}.
These channels have a lot of aspartates and glutamates in their SF. Therefore, at a first glance, they look like a Ca channel.
Hydration plays an important role in the selectivity mechanisms of these channels \cite{2014_finnerty_jpcl_subm}, but this is not the subject of the present study.

Simulation studies for Na channels have been based on models of different resolutions.
All-atom, explicit-water models are usually used when X-ray structures are available.
They are generally studied using molecular dynamics (MD) simulations \cite{2013_ulmschneider_pnasusa_6364,2013_stock_jpcb_3782,2014_finol-urdaneta_jgp_157}.
In the case of the DEKA channel, Lipkind and Fozzard \cite{2008_lipkind_jgp_523} performed MD simulations to explore the Na$^{+}$ vs.\ K$^{+}$ selectivity for various mutants of DEKA based on extreme homology modelling.

Boda et al.\ \cite{boda-pccp-4-5154-2002,boda-bj-93-1960-2007,csanyi-bba-1818-592-2012} and Vora et al.\ \cite{2008_vora_bj_1600} used reduced models of Na channels in the implicit solvent framework.
In these models, only the SF amino acids were represented in an explicit way, while other parts of the channel protein were reduced into a dielectric body.
This is the modelling level that we use in this work.
An intermediate approach is that of Finnerty et al.\ \cite{2013_finnerty_jctc_766}, who proposed a localization method, where SF aminoacid terminal groups are localized into certain positions inspired by structural information.

The advantage of reduced models is that they permit the design of simulation setups in time and length scales that mimic experimental setups and are capable of studying a wide range of concentrations and voltage.
Also, they make it possible to focus on the essential features of the system (SF structure, pore geometry, bath concentrations, voltage, etc.) and to take the effect of the remaining degrees of freedom into account in an averaged, but physically well-based manner (dielectric response as well as external constraints such as the walls of the channel and the membrane).

Simulations can also be distinguished on the basis of the fact whether they were performed in or out of equilibrium.
Equilibrium simulations can study the selective binding of various ions in the SF.
Monte Carlo (MC) simulations, especially in the grand canonical (GC) ensemble (Grand Canonical Monte Carlo, GCMC) are ideal tools for this purpose \cite{boda-pccp-4-5154-2002,boda-bj-93-1960-2007,csanyi-bba-1818-592-2012,2013_finnerty_jctc_766}.
MD simulations can also be used to study selective binding \cite{2008_lipkind_jgp_523,roux_2010,roux_2005}.
Some properties of transport, however, can be extrapolated even from equilibrium simulations on the basis of the integrated Nernst-Planck equation as suggested by Gillespie et al.\ \cite{gillespie-bj-95-2658-2008,boda-jgp-133-497-2009}.
Simulating transport requires a dynamical simulation method.
These can be MD simulations \cite{2013_ulmschneider_pnasusa_6364,2013_stock_jpcb_3782,2014_finol-urdaneta_jgp_157}, Brownian Dynamics simulations \ \cite{2008_vora_bj_1600,chung-bj-77-2517-1999,corry-bj-2001,im_bj_2000}, and Dynamical Monte Carlo (DMC) simulations \cite{boda-jpcc-118-700-2014,csanyi-bba-1818-592-2012,hato-jcp-137-054109-2012}.
Transport can also be studied with theoretical methods such as the Energy Variational approach of Eisenberg et al. \cite{eisenberg_envara,hyon_cms_2011,liu_jpcb_2013,liu_jcp_2014}.

To extend equilibrium binding-selectivity simulations to non-equilibrium situations of steady-state ionic transport is of crucial importance because experimental data are available for ionic currents from electrophysiological measurements \cite{1965_Chandler_jpl_788,1969_binstock_jgp_342,1971_hille_jgp_599,1972_feldman_ab_317,1972_hille_jgp_637,1973_meves_jpl_225,1976_binstock_jgp_551,1976_campbell_jgp_295,1976_ebert_jgp_327,1992_heinemann_n_441,1994_canessa_n_463,1995_tomaselli_bj_1814,2000_sheng_jbc_8572,2001_kellenberger_jgp_679,2003_li_jbc_13867,2007_anantharam_jgp_55}.
The relation of the fluxes carried by the competing ions (flux ratio) defines dynamical selectivity.
How binding selectivity is related to dynamical selectivity is, however, a non-trivial issue as shown by Rutkai et al.\ \cite{rutkai-jpcl-1-2179-2010}.
In particular, the flux is determined not only by the occupancy of a given ionic species in the channel, but also by its mobility.

Measurements show permeability ratios 0.06 and 0.13 for K$^{+}$/Na$^{+}$ and Ca$^{2+}$/Na$^{+}$, respectively \cite{1965_Chandler_jpl_788,1969_binstock_jgp_342,1971_hille_jgp_599,1972_feldman_ab_317,1972_hille_jgp_637,1973_meves_jpl_225,1976_binstock_jgp_551,1976_campbell_jgp_295,1976_ebert_jgp_327,1992_heinemann_n_441}, while $<0.01$ flux ratio for K$^{+}$/Na$^{+}$ \cite{1994_canessa_n_463,1995_tomaselli_bj_1814,2000_sheng_jbc_8572,2001_kellenberger_jgp_679,2003_li_jbc_13867,2007_anantharam_jgp_55}.
To a first approximation, we can assume that binding selectivity agrees well with the above selectivity values measured in terms of flux.
To what degree this assumption is valid can be studied by dynamical simulation methods.
Our first attempt in this direction is the DMC study of Cs\'anyi et al. \cite{csanyi-bba-1818-592-2012}.

In this paper, we focus on equilibrium binding, so it is a direct continuation of our previous papers \cite{boda-pccp-4-5154-2002,boda-bj-93-1960-2007}, where the binding selectivity of the DEKA locus was studied with GCMC simulations using a reduced model of the SF.
These studies used the charge-space competition (CSC) mechanism of Nonner and Eisenberg
\cite{2000_nonner_bj_1976,2001_nonner_jpcb_6427,eisenberg_cpl_2011,crowded_charges,eisenberg_fnl_2012,eisenberg_p_2013,eisenberg_bj_2013}
extended later to inhomogeneous models of the channels studied by GCMC simulations~\cite{boda-jpcb-105-11574-2001,boda-pccp-4-5154-2002,boda-mp-100-2361-2002,boda-jcp-125-034901-2006,boda-prl-98-168102-2007,boda-bj-93-1960-2007,gillespie-bj-95-2658-2008,boda-bj-94-3486-2008,boda-jgp-133-497-2009,malasics-bba-1798-2013-2010}.

The main conclusions of those papers \cite{boda-pccp-4-5154-2002,boda-bj-93-1960-2007} were that K$^{+}$ ions are excluded from the SF by steric repulsion, while Ca$^{2+}$ ions are excluded by an electrostatic penalty.
The new aspect of this study is that we provide an energetic analysis for the phenomena described in our 2007 paper \cite{boda-bj-93-1960-2007}.
The energetic analysis is performed by separating the free energy (more exactly, the chemical potential) into various terms corresponding to various interactions such as volume exclusion, ion-ion, ion-dielectrics, self-energy, etc., interactions.
This approach was introduced by Gillespie \cite{gillespie-bj-2008,dirk-janhavi-mike} in his density functional studies
for the Ryanodine Receptor Ca channel and was extended to three-dimensional models including inhomogeneous
dielectrics using a GCMC methodology \cite{boda-jcp-134-055102-2011} on the basis of Widom's particle insertion
method~\cite{widom63,widom78}.

In our previous work, we analyzed the energetics of the selectivity of the L-type Ca channel \cite{boda-jcp-134-055102-2011}.
In that paper, the dielectric constant of the protein ($\epsilon_{\mathrm{pr}}$) was allowed to be different from that of the baths ($\epsilon_{\mathrm{w}}$).
It was shown that the low dielectric protein surrounding the pore focusing the electric field, and thus enhancing the electrostatics, is necessary to reproduce the micromolar selectivity observed for the L-type Ca channels \cite{AlmersMcCleskey1984,Almersetal1984}.
We also extended that work for the case of a dielectric constant different inside the channel ($\epsilon_{\mathrm{ch}}$) from that of the bath \cite{boda-jcp-139-055103-2013}.
This model is a simple representation of solvation.
Our results showed that solvation plays a minor role in the selectivity mechanism of the L-type Ca channel.
The explanation is that the solvation penalty for Ca$^{2+}$ is balanced by stronger interactions of Ca$^{2+}$ with the SF charges.
Our simulations extending this work to the DEKA locus are in progress and will be published in a subsequent paper.

In this work, however, we restrict ourselves to the case, where the dielectric constants inside and outside the channel are the same ($\epsilon_{\mathrm{ch}}=\epsilon_{\mathrm{w}}$).
This is the model that was considered in our work \cite{boda-bj-93-1960-2007} in 2007.
The SF of the Ca channel is highly charged (EEEE locus, four glutamate residues providing $-4e$ charge).
The DEKA filter, on the other hand, is weakly charged ($-1e$ altogether).
Therefore, it does not favor divalent ions (Ca$^{2+}$).
Additionally, the bulky terminal group of the lysine is present, which, according to our hypothesis, is there to exclude large ions such as K$^{+}$.
This paper examines how these mechanisms work and their energetic basis.

\section{Model}
\label{sec:model}

\begin{figure}[!t]
\begin{center}
{\includegraphics[width=0.47\textwidth]{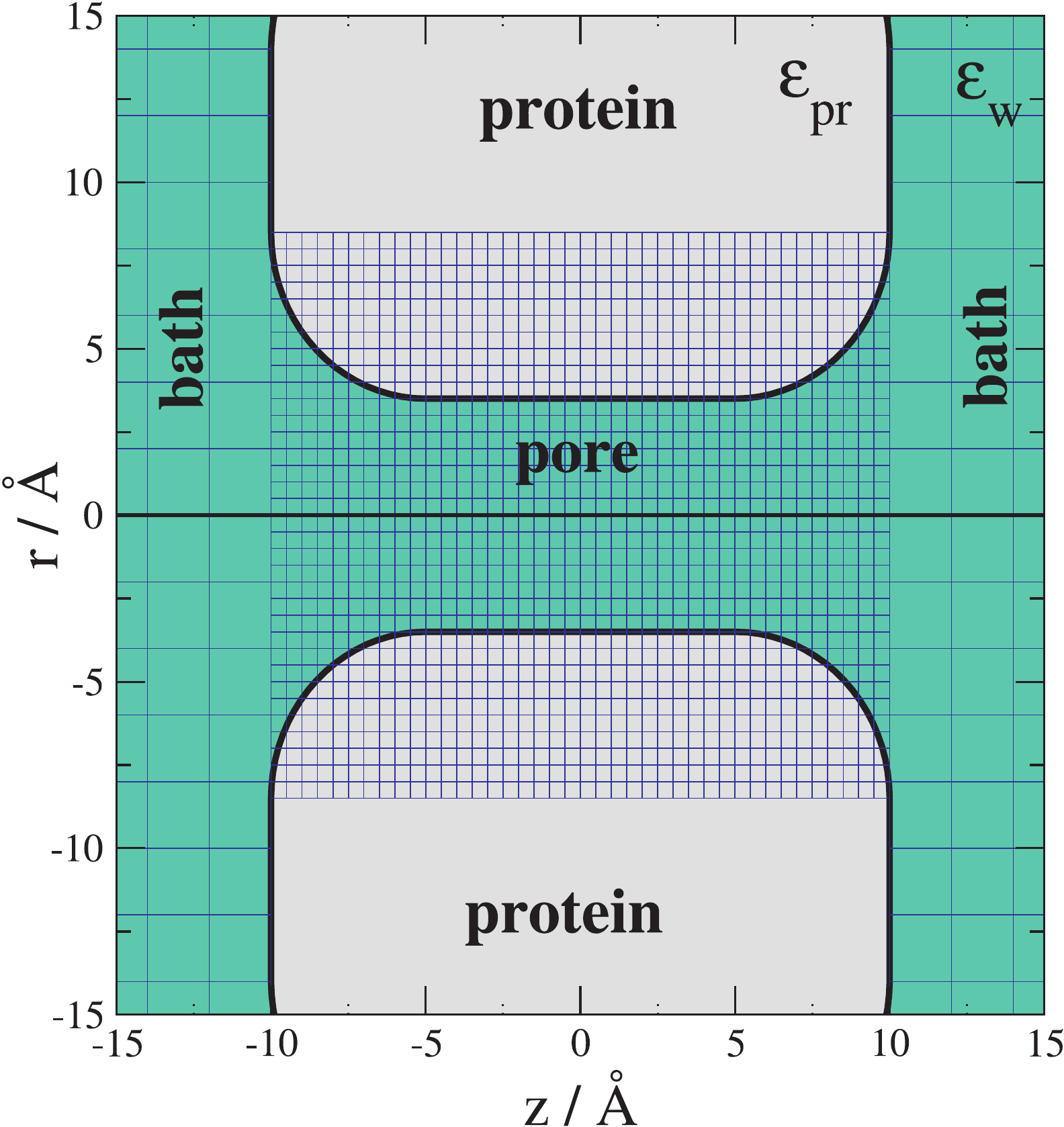}}
\hspace{5mm}
{\includegraphics[width=0.36\textwidth]{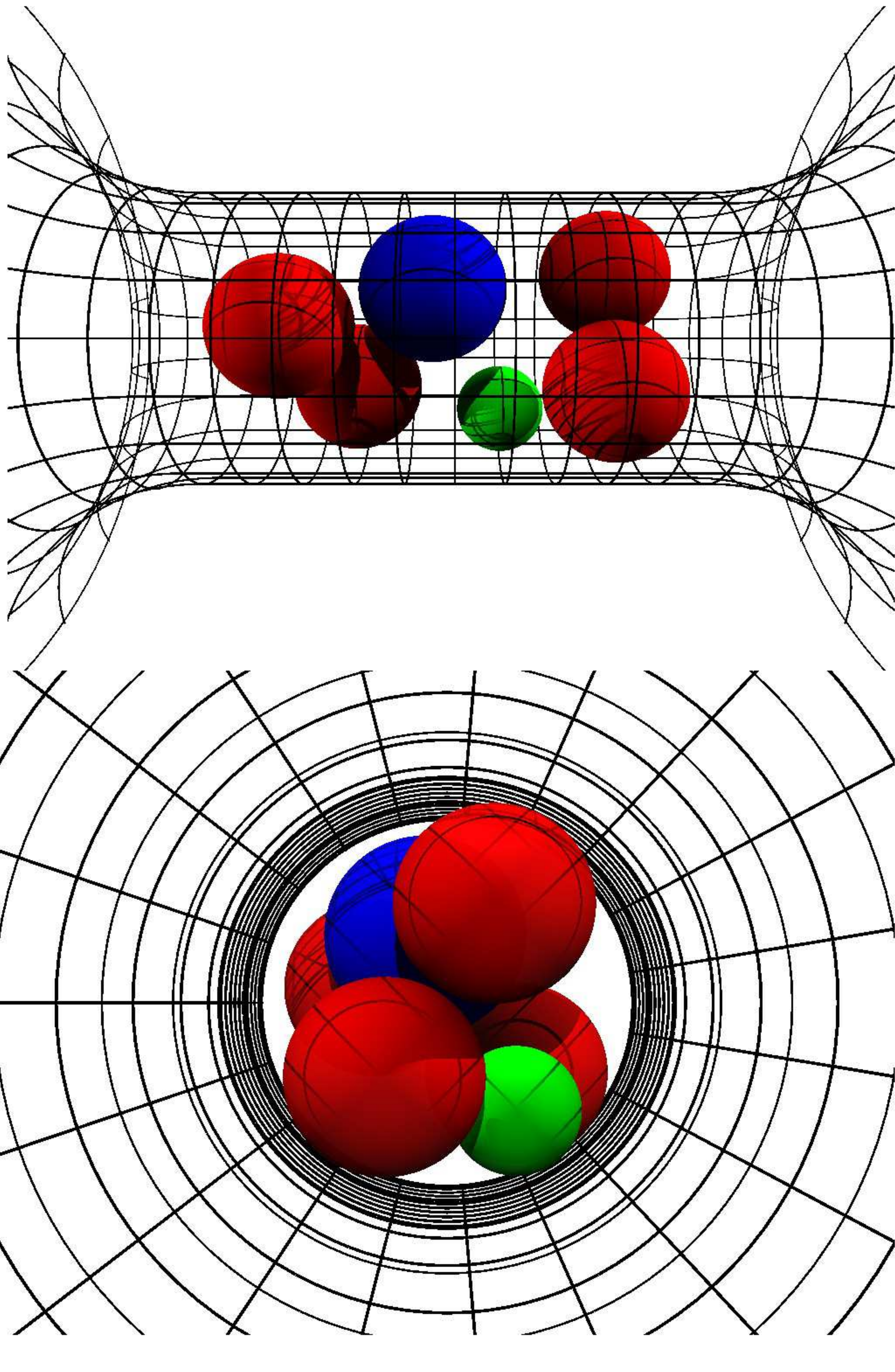}}
\end{center}\caption{(Color online) Model of ion channel, membrane, and electrolyte. The three-dimensional geometry (B) is obtained by rotating the two-dimensional shape shown in panel A around the $z$-axis. The simulation cell is much larger than shown in the figure. The blue lines represent the grid over which the excess chemical potential profiles are computed. The grid is finer inside the channel (width 0.5~{\AA}), while it is coarser outside the channel (width 2~{\AA}).
The selectivity filter ($|z|<5$~{\AA}) contains 4 half charged oxygen ions O$^{1/2-}$ (\textit{red spheres} in panel B) and an ammonium ion NH$_{4}^{+}$ (\textit{blue sphere} in panel B).
For the radii of the ions, the Pauling radii are used: 0.6, 0.95, 1.33, 1.52, 1.7, 0.99, 1.81, 1.4, and 1.5~{\AA} for Li$^{+}$, Na$^{+}$, K$^{+}$, Rb$^{+}$, Cs$^{+}$, Ca$^{2+}$, Cl$^{-}$, O$^{1/2-}$, and NH$_{4}^{+}$ respectively.}
\label{Fig1}
\end{figure}

In our model, most of the atomic structure of the Na channel is reduced to a coarse-grained geometry (figure~\ref{Fig1}).
The channel protein is represented as a continuum solid with dielectric coefficient $\epsilon_{\mathrm{pr}}$.
The three dimensional body of the protein is obtained by rotating the thick line in figure~\ref{Fig1} about the $r=0$ axis.
The protein thus forms an aqueous pore that connects the two baths.
Water in the baths and pore is described as an implicit solvent that is a continuum dielectric having uniform dielectric coefficient $\epsilon_{\mathrm{w}}=80$.
The central cylindrical part of the pore (with radius $R=3.5-4.5$~{\AA} and length 10~{\AA})  forms the selectivity filter that includes the only atoms of the protein that are treated explicitly.
These atoms are four half-charged `oxygen ions' O$^{1/2-}$ [figure~\ref{Fig1}~(B), \emph{red spheres}] representing the charged terminal groups of the D and E residues, while a positive `ammonium ion' NH$_{4}^{+}$ [figure~\ref{Fig1}~(B), \emph{blue sphere}] represents the terminal group of the K residue.
The alanine is ignored.
The structural oxygen ions are confined to the selectivity filter (their centers are in the region $r \leqslant R-R_{i}$, $|z| \leqslant 5~\text{\AA}-R_{i}$, where $R_{i}$ is the ionic radius), but they can move freely inside the filter.

The ions are modelled as charged hard spheres with crystal radii (see caption of figure~\ref{Fig1}).
The computation of the intermolecular energy terms due to screened Coulomb potentials and interactions with polarization charges induced on the dielectric boundaries [the boundary of the protein and the electrolyte; thick line in figure~\ref{Fig1}~(A)] are described in our previous works \cite{boda-pre-69-046702-2004,boda-jcp-125-034901-2006,boda-jcp-134-055102-2011}.
Ions are restricted to the aqueous space of the model and cannot overlap with hard walls in the system.
Figure~\ref{Fig1}~(A) shows only the small central region of the simulation cell.
The entire simulation cell is a cylinder having typical dimensions of radius 40~{\AA} and length 180~{\AA}.
The channel is embedded in a membrane region that excludes ions by hard walls as described  before \cite{boda-jcp-125-034901-2006}.

\section{Method of energetic analysis}
\label{sec:energetic}

In an equilibrium GCMC simulation, the acceptance of ion insertion/deletions of ions is governed by the configurational chemical potential of the respective ionic species $i$ defined as
\begin{equation}
\mu_{i} = kT \ln c_{i} (\mathbf{r}) + \mu^{\mathrm{EX}}_{i} (\mathbf{r}) = kT\ln c_{i}(\mathrm{B}) + \mu_{i}^{\mathrm{EX}}(\mathrm{B}),
\label{eq:chempot}
\end{equation}
where $k$ is Boltzmann's constant, $T$ is the temperature, $c_{i}(\mathbf{r})$ is the concentration profile, $\mu_{i}^{\mathrm{EX}}(\mathbf{r})$ is the excess chemical potential profile, $c_{i}(\mathrm{B})$ is the bulk concentration, and  $\mu_{i}^{\mathrm{EX}}(\mathrm{B})$ is the bulk excess chemical potential.
Although $kTc_{i}(\mathbf{r})$ and $\mu_{i}^{\mathrm{EX}}(\mathbf{r})$ can be different in different regions (they are position dependent), their sum is constant due to equilibrium.
The bulk excess chemical potentials $\mu_{i}^{\mathrm{EX}}(\mathrm{B})$ corresponding to the prescribed bulk concentrations $c_{i}(\mathrm{B})$ are calculated using the Adaptive GCMC method \cite{malasics-jcp-132-244103-2010}.
By rewriting equation (\ref{eq:chempot}), the excess chemical potential difference is defined as
\begin{equation}
\Delta \mu_{i}^{\mathrm{EX}}(\mathbf{r}) = \mu_{i}^{\mathrm{EX}}(\mathbf{r})-\mu_{i}^{\mathrm{EX}}(\mathrm{B}) = -kT \ln \left(\dfrac{c_{i}(\mathbf{r})}{c_{i}(\mathrm{B})} \right) .
\label{eq:deltamu_ex}
\end{equation}
It can be identified with the binding free energy of an ion moved from a bath (B) to position $\mathbf{r}$ of the channel \cite{boda-jcp-134-055102-2011}.
If we write up equation (\ref{eq:deltamu_ex}) for Na$^{+}$ and K$^{+}$ and take the difference, we can derive that
 \begin{equation}
\ln \left( \dfrac{c_{\mathrm{Na}^{+}}(\mathbf{r})}{c_{\mathrm{K}^{+}}(\mathbf{r})} \right)  =
\ln \left( \dfrac{c_{\mathrm{Na}^{+}}(\mathrm{B})}{c_{\mathrm{K}^{+}}(\mathrm{B})} \right) +
\dfrac{\Delta\Delta \mu^{\mathrm{EX}}(\mathbf{r})}{kT} \,,
\label{eq:finaladv}
\end{equation}
where
\begin{equation}
\Delta\Delta\mu^{\mathrm{EX}} (\mathbf{r}) = \Delta \mu^{\mathrm{EX}}_{\mathrm{Na}^{+}} (\mathbf{r})  -  \Delta \mu^{\mathrm{EX}}_{\mathrm{K}^{+}} (\mathbf{r})   .
\end{equation}
Similar equations can be given for other pairs of ions.

In equation~(\ref{eq:finaladv}), the left-hand side is called `binding selectivity' because it expresses the degree to which Na$^{+}$ is favored over K$^{+}$ at location $\mathbf{r}$ (binding selectivity is positive if location $\mathbf{r}$ is selective for Na$^{+}$ over K$^{+}$).
The corresponding term on the right-hand side containing the bulk concentrations is called `number advantage' \cite{gillespie-bj-2008} because it expresses the advantage that an ionic species gets from outnumbering the other species in the bulk.
The channel can become selective for a given ionic species for two reasons: either from the number advantage or the energetic advantage expressed by $\Delta \Delta \mu^{\mathrm{EX}} (\mathbf{r})$.

The energetic advantage, however, contains terms due to different interactions present in the system as described in appendix~\ref{sec:wid}.
The EX term can be divided in various ways.
Here, we use the division used in our latest work \cite{boda-jcp-139-055103-2013}:
\begin{equation}
\Delta\mu^{\mathrm{EX}}_{i}(\mathbf{r}) = \Delta\mu^{\mathrm{HS}}_{i}(\mathbf{r}) + \Delta\mu^{\mathrm{II}}_{i}(\mathbf{r})  + \Delta\mu^{\mathrm{ID}}_{i}(\mathbf{r}) + \Delta\mu^{\mathrm{SELF}}_{i}(\mathbf{r})
\end{equation}
or briefly $\mathrm{EX}=\mathrm{HS}+\mathrm{II}+\mathrm{ID}+\mathrm{SELF}$,
where HS means hard sphere exclusion, II means interaction with the ions, ID means interactions with the dielectrics (polarization charges induced by other ions), and SELF means interactions with the polarization charges induced by the ion itself.
(In the division of our earlier work \cite{boda-jcp-134-055102-2011}, we used the DIEL term that contained the SELF term, namely, $\mathrm{DIEL}=\mathrm{ID}+\mathrm{SELF}$.)
We can also use the division $\mathrm{EX}=\mathrm{HS}+\mathrm{MF}+\mathrm{SC}+\mathrm{SELF}$, where MF means the interaction with the mean (average) electric field of all the existing charges in the system (ionic and induced).
SC expresses correlations beyond the mean field level (SC refers to `screening') \cite{gillespie-bj-2008}.
The SELF term is a one-particle term (mean-field in nature) and corresponds to the average electrostatic interaction energy of the inserted ion with its self-induced charge.
It is not included in the ID or the MF term.
The SELF term corresponds to the dielectric boundary force or energy of reference~\cite{nadler_2003}.

The computation of all these terms can be found in our original paper \cite{boda-jcp-134-055102-2011} and in appendix~\ref{sec:wid}.
Briefly, the total EX chemical potential can unambiguously be obtained by inserting charged hard spheres (representing the ions) in the Widom particle insertion method.
Different terms of EX are computed by inserting particles interacting only through short-ranged (HS) or more direct (II) interactions and obtaining the rest as residuals.
For example, it is reasonable to compute the HS term by inserting uncharged hard spheres with the same radius as the respective ion in the Widom procedure.
All the remaining terms (II, ID, SELF) are electrostatic in nature and obtained by deducting the HS term from the EX term.
(The separation of HS and electrostatic terms and their effect on selectivity can already be found in the work of Nonner et al. \cite{2000_nonner_bj_1976} in the context of the mean spherical approximation.)
Similar procedures are used to separate the II and ID, as well as the MF and SC terms, as described in appendix~\ref{sec:wid}.

The $\mathbf{r}$-dependence of various terms is computed by ion insertions into grid cells shown in figure~\ref{Fig1}.
Note that the concentration profile can be computed in two different ways.
First, sampling the number of ions in a volume element, computing the average ion number and dividing by the volume of the element.
This is advantageous when the concentration and/or the volume element is large so there is a large enough sample of ions.
The concentration, on the other hand, can also be computed from equation (\ref{eq:chempot}) by computing the EX term from the Widom method and deducting it from the chemical potential.
This approach is useful where the concentration is low.
This method was used in our simulations for the DEKA channel.

Our grid is two-dimensional because we have rotational symmetry.
Our profiles, therefore, are expressed in terms of the $(z,r)$ cylindrical coordinates.
In this work, however, we show the results that are averaged over the $r$-coordinate
\begin{equation}
\Delta \mu_{i}^{\mathrm{EX}}(z) = \dfrac{2}{R_{\mathrm{min}}^{2}(z)} \int_{0}^{R_{\mathrm{min}}(z)} r\, \Delta \mu_{i}^{\mathrm{EX}} (z,r) \,\rd r ,
\label{r-average}
\end{equation}
where $R_{\mathrm{min}}(z)=R(z)-R_{\mathrm{larger \, \, ion}}(z)$ is the cross-section that is accessible to the center of the larger of the competing ions, [$R(z)$ denotes the radius of the simulation domain at $z$].

\section{Results and discussion}
\label{sec:results}

We start our discussion with competition of ions of the same charge.
Specifically, we study the selectivity of Na$^{+}$ over various monovalent ions.
In the classical mole fraction experiment, the mole fraction of one ion (Na$^{+}$, for example) is changed while keeping the total cation concentration constant (when divalent ion is present, the total ionic strength is kept constant in some studies).
These results are seen in figure~5 of reference \cite{boda-bj-93-1960-2007}.
In this work, the concentration of the two competing cations in the baths is the same (50~mM), so the number advantage is zero.

\begin{figure}[!t]
\begin{center}
\hspace{10mm}{\includegraphics[height=0.75\textwidth,angle=-90]{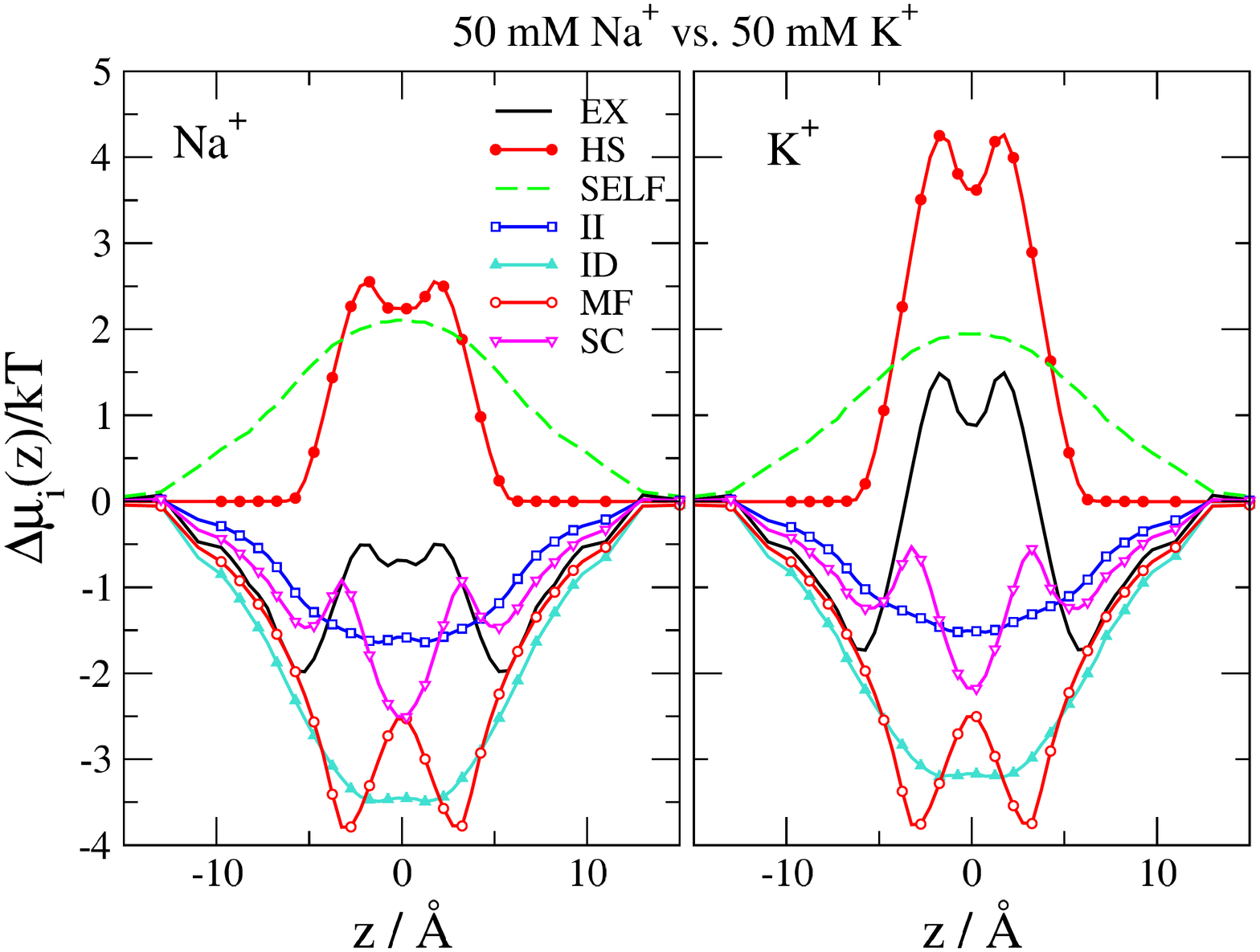}}
\end{center}
\vspace{-2mm}\caption{(Color online) The $\Delta \mu_{i}(z)$-profiles for Na$^{+}$ and K$^{+}$ for the case when the bath concentration is the same for the two competing monovalent cations (50~mM) for $\epsilon_{\mathrm{pr}}=10$ and $R=3.5$~{\AA}.}
\label{Fig2}
\end{figure}

Figure~\ref{Fig2} shows the various terms of the $\Delta \mu ^{\mathrm{EX}}_{i}(z)$-profiles for Na$^{+}$ and K$^{+}$ for protein dielectric constant $\epsilon_{\mathrm{pr}}=10$ and filter radius $R=3.5$~{\AA}.
The value $\epsilon_{\mathrm{pr}}=10$ is the value fixed in our studies for the L-type Ca channel \cite{boda-jcp-125-034901-2006,boda-prl-98-168102-2007,gillespie-bj-95-2658-2008,boda-jgp-133-497-2009,malasics-bba-1788-2471-2009,malasics-bba-1798-2013-2010,boda-jcp-134-055102-2011,giri-pb-8-026004-2011}.
The value $R=3.5$~{\AA} was used in our DMC study for the DEKA Na channel to reproduce experimental data \cite{csanyi-bba-1818-592-2012}.

The EX terms are related to the concentration ratios through $-\ln [c_{i}(\mathbf{r})/c_{i}(\mathrm{B})]$ [see equation~(\ref{eq:deltamu_ex})].
Therefore, where the EX term (or any component) is negative, it energetically favors the ionic species, so it increases the concentration of that ionic species.
As also seen in figure~6 of our previous paper \cite{boda-bj-93-1960-2007}, there are peaks at the entrances of the SF and the vestibules ($|z|\sim 5$~{\AA}).
In the center of the SF, on the other hand, the concentrations are low.
This region forms a depletion zone for both ions, where ions have difficulty to enter.
The question, therefore,  is which ion is excluded less from this region.
The answer is that there are more Na$^{+}$ than K$^{+}$ in the SF (the EX term is lower for Na$^{+}$), so the SF is Na$^{+}$-selective.

All the electrostatic terms (II, ID, MF, SC) are negative except the SELF term.
The SELF term is repulsive because the ions are in the $\epsilon_{\mathrm{w}}=80$ region, so the sign of the induced charge on the $\epsilon_{\mathrm{pr}}|\epsilon_{\mathrm{w}}$ boundary is the same as the sign of the inserted ion itself.
This practically corresponds to the dielectric penalty an ion must pay when it passes the low dielectric membrane region as described in classical works \cite{neumcke_bj_1969, parsegian_n_1969,levitt_bj_1978,jordan_bj_1982}.
The SELF term is slightly larger for Na$^{+}$ because the smaller Na$^{+}$ can get closer to the channel wall and can induce a larger polarization charge.

The other term that is positive is the HS term describing the volume exclusion.
This is the term that is very different in the case of Na$^{+}$ and K$^{+}$; it is larger in the case of K$^{+}$.
Since the size of K$^{+}$ ions (we talk about the dehydrated (Pauling) radius) is larger, it is more difficult to insert such an ion in the SF.
Therefore, K$^{+}$ has a larger entropic penalty than Na$^{+}$ does.
This difference is especially apparent in the center of the SF, where the NH$_{4}^{+}$  (the structural ion representing the large terminal group of the lysine)  profile has a peak (see figure~6 of reference \cite{boda-bj-93-1960-2007}).
Without the HS term (ions of finite size) we could not get a Na$^{+}$-selective filter (against K$^{+}$) in this model.

The MF term is negative, because the SF is negatively charged.
There is no space for the cations to fully neutralize the SF charge.
The SC term is similar to the MF term in order of magnitude indicating that mean field theories are not sufficient to study ionic systems in crowded confined spaces such as the SF of ion channels.

\begin{figure}[!t]
\begin{center}
{\includegraphics[width=0.85\textwidth]{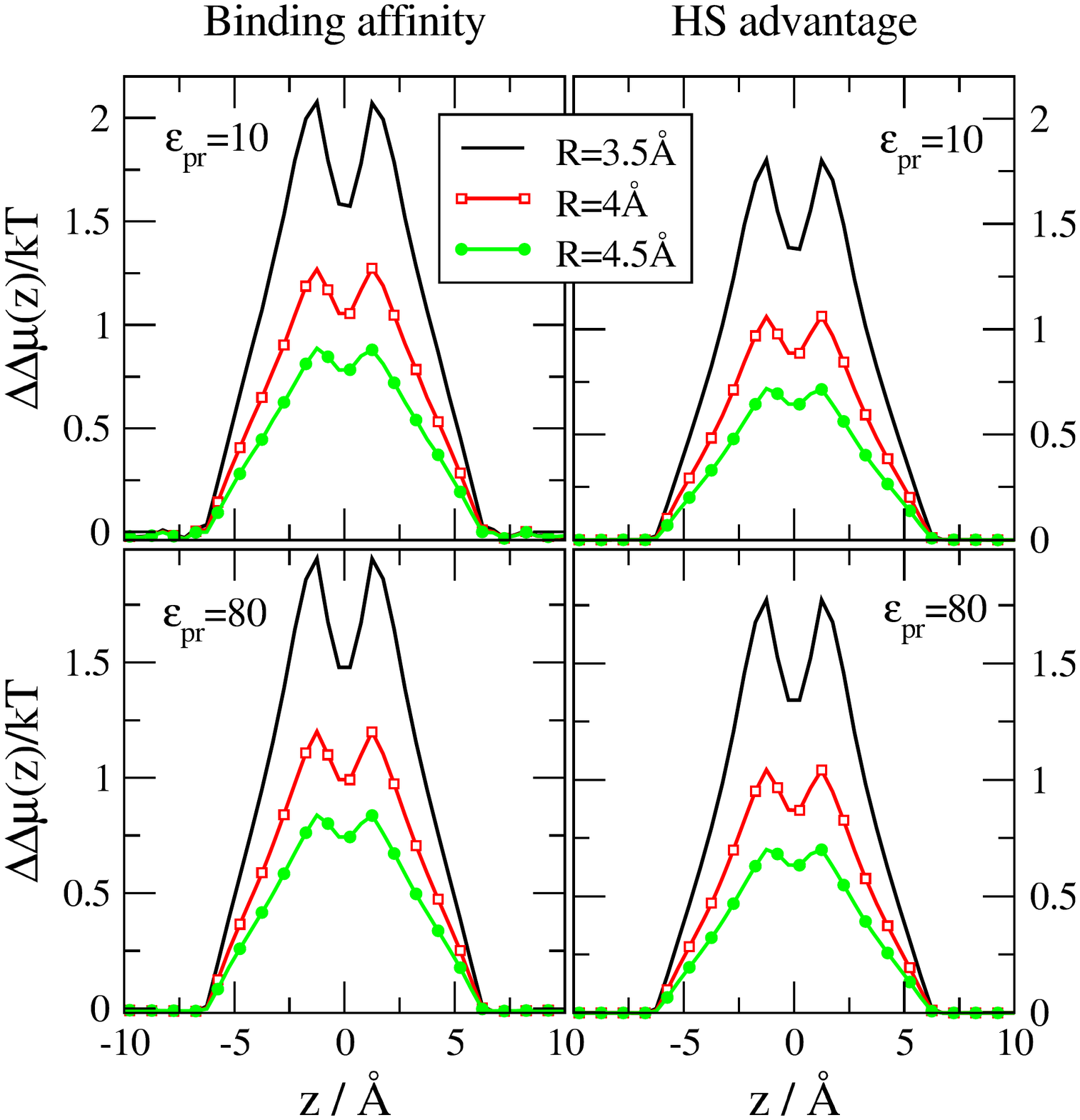}}
\end{center}
\vspace{-2mm}
\caption{ (Color online)
The binding affinity and HS advantage curves for Na$^{+}$ vs.\ K$^{+}$ competition for $\epsilon_{\mathrm{pr}}=10$ (top panels) and 80 (bottom panels) for filter radii $R=3.5$, 4, and 4.5~{\AA} (50--50~mM bath concentrations).}
\label{Fig3}
\end{figure}

\begin{figure}[!b]
\begin{center}
{\includegraphics[width=0.85\textwidth]{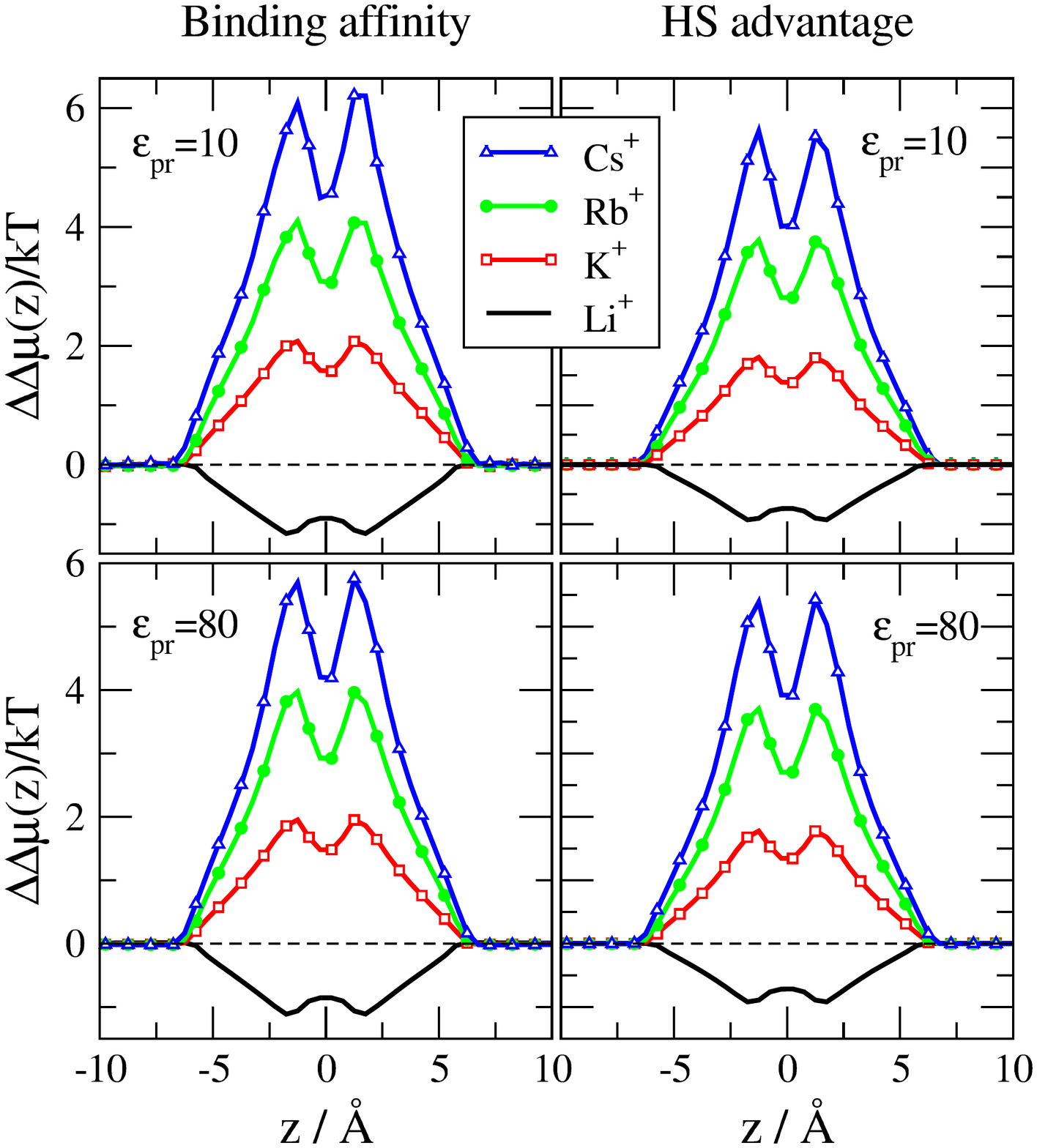}}
\end{center}
\vspace{-2mm}
\caption{(Color online) The binding affinity and HS advantage curves for the competition of Na$^{+}$ against various monovalent ions (Li$^{+}$, K$^{+}$, Rb$^{+}$, Cs$^{+}$) for $\epsilon_{\mathrm{pr}}=10$ (top panels) and 80 (bottom panels) (50--50~mM bath concentrations, $R=3.5$~{\AA}).
}
\label{Fig4}
\end{figure}

The dominant term that drives Na$^{+}$ vs.\ K$^{+}$ selectivity is the HS term.
In figures~\ref{Fig3} and \ref{Fig4}, therefore, only the differences of the EX and HS terms are shown for various cases.
In this special case, where the number advantage is zero, the EX difference is equal to the binding affinity [see equation (\ref{eq:finaladv})], while the HS term is the dominant term of $\Delta\Delta\mu^{\mathrm{EX}}(\mathbf{r})$.
Since the differences are obtained by deducting the K$^{+}$ terms from the Na$^{+}$ terms, positive values favor Na$^{+}$.


Figure \ref{Fig3} shows the profiles for various pore radii for $\epsilon_{\mathrm{pr}}=10$ (top panels) and   $\epsilon_{\mathrm{pr}}=80$ (bottom panels).
Narrower channels favor Na$^{+}$ even more, as expected, because it is even more difficult to find space for the large K$^{+}$ ions in the small SF compared to Na$^{+}$.
Putting it in another way, Na$^{+}$ vs.\ K$^{+}$ selectivity is better for narrow channels, where stronger competition is forced by the confinement and lack of space, so the smaller size of Na$^{+}$ has the advantage.
The binding affinity curves (left-hand panels) and the HS advantages (right-hand panels) behave similarly with small differences due to other energetic terms (see figure~\ref{Fig2}).

Another conclusion of the figure is that Na$^{+}$ vs.\ K$^{+}$ selectivity does not depend on the dielectric constant of the protein; the curves for $\epsilon_{\mathrm{pr}}=10$ (top panels) and  $\epsilon_{\mathrm{pr}}=80$ (bottom panels) behave practically the same.

Figure~\ref{Fig4} shows the same curves but now for a fixed pore radius ($R=3.5$~{\AA}) and different monovalent cations (Li$^{+}$, K$^{+}$, Rb$^{+}$, Cs$^{+}$) competing with Na$^{+}$.
The main conclusion is similar to those drawn in figure~\ref{Fig3}; the crowded SF favors the smaller ion.
The pore is selective for Li$^{+}$ against Na$^{+}$, while it is selective for Na$^{+}$ against the larger ions.

The protein dielectric constant does not have an effect on these profiles.
Of course, the value of $\epsilon_{\mathrm{pr}}$ has a large effect on the individual ionic profiles and the occupancies (see figure~8 of reference \cite{boda-bj-93-1960-2007}), but not on the relative ones that we study here.

In the second half of this section, we analyze the competition of Na$^{+}$ against Ca$^{2+}$.
The other usual way to study the behavior of the channel having varying electrolyte composition is to keep the concentration of one species fixed (Na$^{+}$, for example) and to add another species (Ca$^{2+}$, for example) gradually.
This added salt experiment was done by Almers and McCleskey in their experiment for the L-type Ca channel \cite{Almersetal1984,AlmersMcCleskey1984}.
We performed this kind of experiment in our previous simulations for the DEKA locus and its DEEA mutant, see figure~2 of reference~\cite{boda-bj-93-1960-2007}.

\begin{figure}[!b]
\begin{center}
{\includegraphics[width=0.9\textwidth]{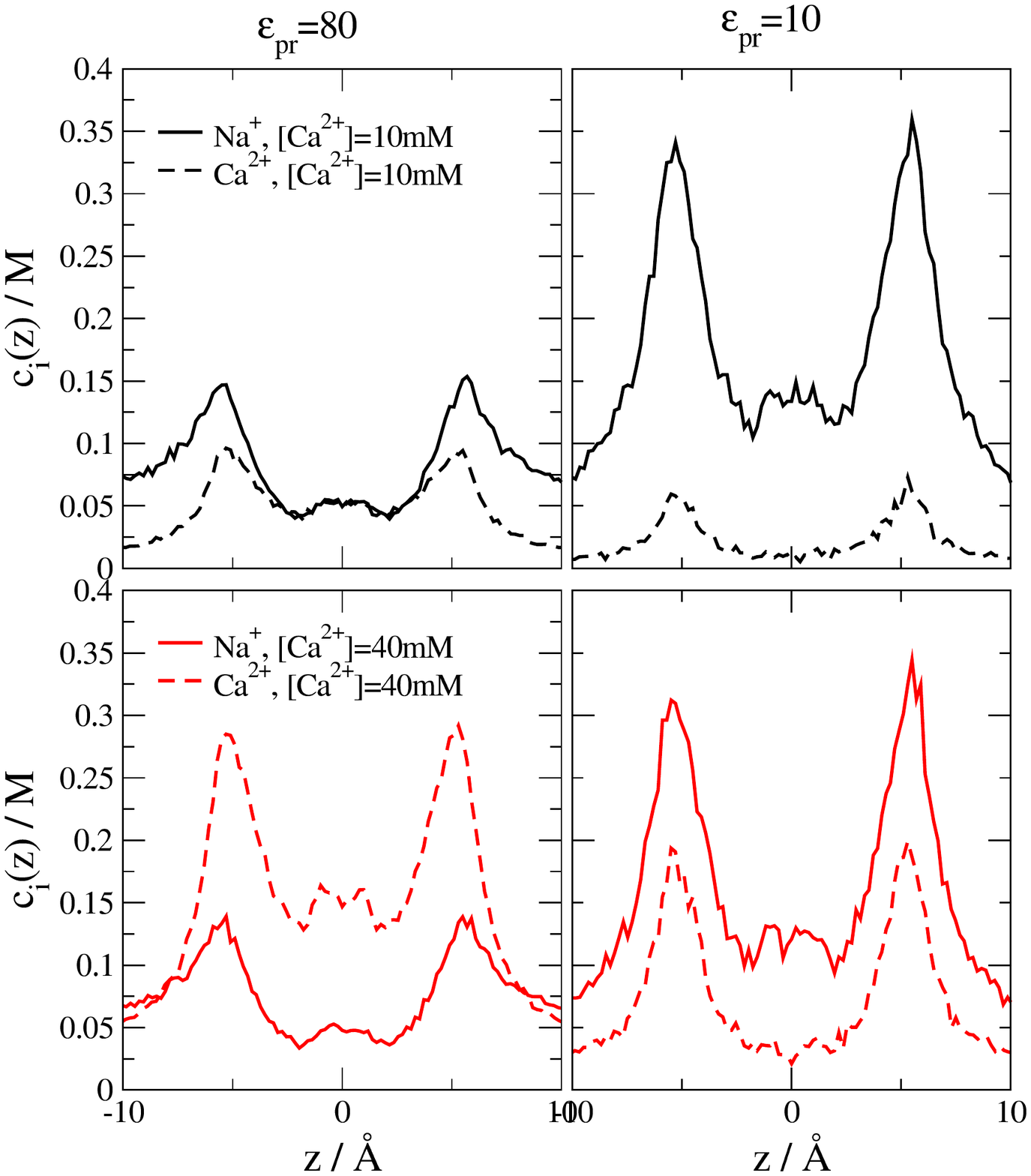}}
\end{center}
\vspace{-4mm}
\caption{(Color online) Ca$^{2+}$ and \ Na$^{+}$ concentration profiles for two different Ca$^{2+}$ concentrations (10 and 40~mM in top and bottom panels, respectively) with a 50~mM Na$^{+}$ background ($R=3.5$~{\AA}). The profiles are shown for protein dielectric constants $\epsilon_{\mathrm{pr}}=80$ (left-hand panels) and 10 (right-hand panels).}
\label{Fig5}
\end{figure}

Those simulations qualitatively reproduced the experiment of Heinemann et al. \cite{1992_heinemann_n_441}.
Heinemann et al. found that mutating the DEKA locus into a DEEA locus, the selectivity behavior of the channel is reminiscent to Ca channels rather than Na channels.
In experiment, the current drops to half (IC50) at Ca$^{2+}$ concentration $10^{-4}$~M, while in our simulations, the number of Na$^{+}$ ions drops to half at the same concentration.
The explanation is that the DEEA mutation has $-3e$ charge producing a Ca channel, but with weaker selectivity than in the case of the $-4e$ charge (EEEE locus).
The DEKA locus, on the other hand, shows Na$^{+}$ over Ca$^{2+}$ selectivity.
This selectivity is stronger for smaller $\epsilon_{\mathrm{pr}}$ (see figure~10~(A) of reference~\cite{boda-bj-93-1960-2007}).
The dielectric constant of the protein, therefore, has a strong effect in the case of monovalent vs. divalent competition.

In figure~\ref{Fig5}, we show the results only for two chosen Ca$^{2+}$ concentrations, 10~mM (top panels) and 40~mM (bottom panels)~--- both are well above the physiological values ($\sim$1--2~mM).

The background Na$^{+}$ concentration is 50~mM.
The Na$^{+}$ and Ca$^{2+}$ concentration profiles are shown for $\epsilon_{\mathrm{pr}}=80$ (left-hand panels) and 10 (right-hand panels).

There are more Na$^{+}$ than Ca$^{2+}$ ions in the filter in the case of $\epsilon_{\mathrm{pr}}=10$ for both concentrations.
A single Na$^{+}$ ion efficiently counterbalances the filter charge.
Ca$^{2+}$ ions, on the other hand, overcharge the filter, which is electrostatically unfavorable.
To counterbalance this overcharge, a Cl$^{-}$ would be needed, but there is no space left for it in the filter.

In the case $\epsilon_{\mathrm{pr}}=80$, on the other hand, there are more Ca$^{2+}$ ions at [Ca$^{2+}]=40$~mM.
The explanation is that Ca$^{2+}$ is still double charged so the SF attracts it more strongly.
The overcharged filter is balanced by Cl$^{-}$ ions from outside the filter.
In this case, this is possible because the Coulomb forces are more long-ranged and more screened than in the case of $\epsilon_{\mathrm{pr}}=10$, where the low-dielectric protein focuses the electric field.
This means that the low dielectric protein is needed to exclude Ca$^{2+}$.

\begin{figure}[!b]
\begin{center}
{\includegraphics[height=0.75\textwidth,angle=-90]{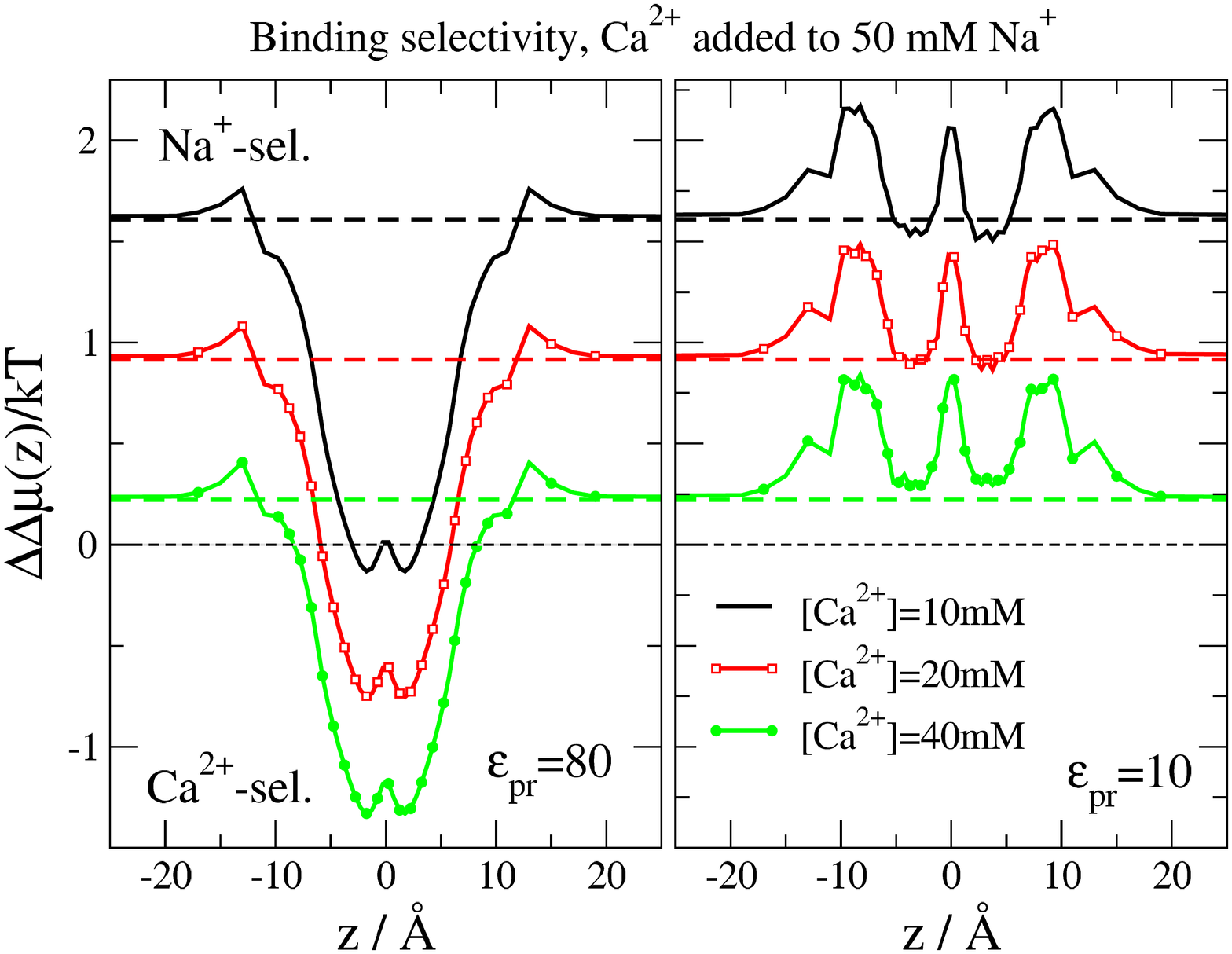}}
\end{center}
\vspace{-2mm}
\caption{(Color online) Binding selectivity (solid lines) and number advantage (dashed lines)
curves for the cases considered in figure~\ref{Fig5} for three different concentrations.}
\label{Fig6}
\end{figure}

\begin{figure}[!t]
\begin{center}
{\includegraphics[height=0.85\textwidth,angle=-90]{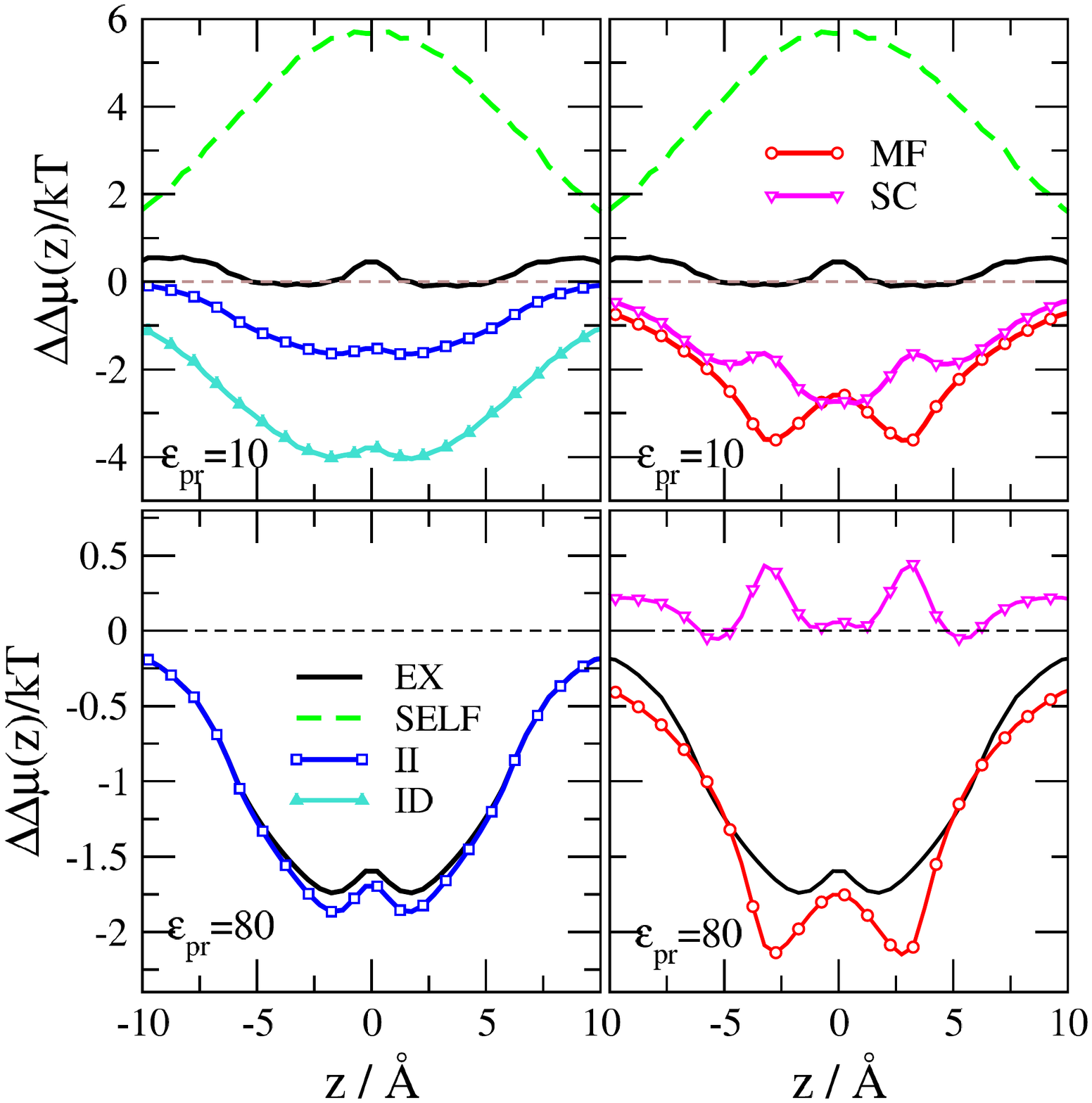}}
\end{center}
\vspace{-2mm}
\caption{(Color online) The various terms of the free energy advantage for [Ca$^{2+}$]=10~mM and [Na$^{+}$]=50~mM.
Top and bottom panels show the results for $\epsilon_{\mathrm{pr}}=10$ and 80, respectively.
Left panels show the EX curve in the $\mathrm{EX} = \mathrm{HS} + \mathrm{II} + \mathrm{ID} + \mathrm{SELF}$ division, while the right panels show the EX curve in the $\mathrm{EX} = \mathrm{HS} + \mathrm{MF} + \mathrm{SC} + \mathrm{SELF}$ division.
}
\label{Fig7}
\end{figure}

The energetics of this phenomenon is analyzed in figures~\ref{Fig6} and \ref{Fig7}.
The difference in Na$^{+}$ vs.\ Ca$^{2+}$ selectivity is more clearly seen by plotting the binding selectivity curves.
When this is positive, the pore is Na$^{+}$-selective, while it is Ca$^{2+}$-selective in the opposite case.
The number advantages are also indicated with dashed horizontal lines.
As bath Ca$^{2+}$ concentration is increased, this line and the binding selectivity curve with it are shifted downwards.
The shape of the binding selectivity curves does not change much with the bath Ca$^{2+}$ concentration.
We can conclude, therefore, that Na$^{+}$ vs. Ca$^{2+}$ selectivity does not depend on the bath Ca$^{2+}$ concentration.
This is because the DEKA locus is a singly occupied SF; only one cation occupies the SF at one time (or none).

This was not true for the L-type Ca channel.
That channel could be multiply occupied, so selectivity behavior was a function of Ca$^{2+}$ concentration due to correlations of cations in the filter.
Furthermore, the SF of the EEEE locus became more charge neutral as Ca$^{2+}$ concentration was increased.
Because of that, the MF terms decreased (see figure~7 of Boda et al.~\cite{boda-jcp-134-055102-2011}).
That effect is absent here; the probability that a channel becomes charge neutral does not depend on ionic concentrations, but it rather depends on entropic effects (available space in the channel given by filter radius and ion sizes).

The difference of binding selectivity and number advantage defines the free energy advantage, \linebreak $\Delta\Delta\mu^{\mathrm{EX}} (\mathbf{r})$, [see equation (\ref{eq:finaladv})].
The terms of that advantage are analyzed for $\epsilon_{\mathrm{pr}}=10$ and 80 for a given Ca$^{2+}$ concentration (10~mM) in figure~\ref{Fig7}.

The top panels show the $\epsilon_{\mathrm{pr}}=10$ data.
The left-hand panel shows the II and ID terms ($\mathrm{EX} = \mathrm{HS} + \mathrm{II} + \mathrm{ID} + \mathrm{SELF}$), while  the right-hand panel shows the MF and SC terms ($\mathrm{EX} = \mathrm{HS} + \mathrm{MF} + \mathrm{SC} + \mathrm{SELF}$).
The EX and SELF terms are shown both in the left-hand and right-hand sides.
The HS term (not shown) is close to zero because the ions are of similar sizes.
The EX term is also close to zero in this case, but this is the effect of the balance of the different free energy advantage terms.
The SELF term is very positive, so it favors Na$^{+}$.
This term is about four times larger for Ca$^{2+}$ than for Na$^{+}$ so it plays the role of solvation penalty in this model.
Without the SELF term, we could not get a Na$^{+}$ selective filter (against Ca$^{2+}$) in this model.
Both the II and ID terms (as well as the MF and SC terms, see right-hand panel) favor Ca$^{2+}$ because Ca$^{2+}$ is attracted twice as strongly by the SF charges (ionic and induced) as Na$^{+}$.

The bottom panels show the $\epsilon_{\mathrm{pr}}=80$ data.
Here, the ID and SELF terms are absent, because there is no dielectric boundary present.
The ID term favors Ca$^{2+}$, while the SELF term favors Na$^{+}$.
Since the SELF term is larger in absolute value, these two terms together ($\mathrm{ID}+\mathrm{SELF}$) still favor Na$^{+}$, so the channel becomes less Na$^{+}$ selective in their absence.

The SC term is small for $\epsilon_{\mathrm{pr}}=80$, which means that Na$^{+}$ vs.\ Ca$^{2+}$ selectivity is chiefly a mean-field effect in this case; the O$^{1/2-}$ ions attract Ca$^{2+}$ twice as strongly as they attract Na$^{+}$.
In the case of $\epsilon_{\mathrm{pr}}=10$, on the other hand, SC is quite large indicating a SF of higher density and correlations beyond the mean-field level (mainly, with induced charges).

Summarizing, the EX term is negative for $\epsilon_{\mathrm{pr}}=80$, so it is rather a Ca channel.
The EX term is close to zero for $\epsilon_{\mathrm{pr}}=10$, which means that neither ions are favored energetically.
Binding selectivity is driven by the number advantage, which results in a Na$^{+}$ selective channel (against Ca$^{2+}$) at physiological Ca$^{2+}$ concentrations (1--2~mM).

\clearpage

\section{Conclusions}
\label{sec:concl}

We analyzed the energetics of ion selectivity in the SF of the DEKA Na channels.
The reduced model studied before \cite{boda-bj-93-1960-2007} was capable of reproducing the basic characteristics of this channel.
We showed that K$^{+}$ ions are excluded from the SF due to entropic hard sphere exclusion.
The dielectric constant of the protein has no effect on this selectivity.
In general, this filter favors smaller ions over larger ones.

Ca$^{2+}$ ions, on the other hand, are excluded from the filter due to a free-energetic penalty which is enhanced by the low dielectric constant of the protein.
The DEKA locus works as a Na channel in the Na$^{+}$ vs.\ Ca$^{2+}$ competition by \textit{not favoring} Ca$^{2+}$.
The dominant term is the number advantage in the bulk solutions.
In physiological situations this mechanism suffices.

We showed that the dominant term of the energetic penalty is the SELF term, which is a dielectric penalty~--- the interaction of the ion with the polarization charges induced by itself.
This dielectric penalty is a simple, implicit representation of solvation penalty in the framework of this model, where $\epsilon_{\mathrm{ch}}=\epsilon_{\mathrm{w}}$.
Simulations, where a different dielectric constant inside the channel is used, take solvation into account explicitly.

\section*{Acknowledgements}

We gratefully acknowledge the computing resources provided on Blues and/or Fusion, high-perfor\-mance computing cluster operated by the Laboratory Computing Resource Center at Argonne National Laboratory.
We acknowledge the support of the Hungarian National Research Fund (OTKA NN113527) in the framework of ERA Chemistry.
Present publication was realized with the support of the projects T\'AMOP-4.2.2/A-11/1/KONV-2012-0071 and T\'AMOP-4.1.1/C-12/1/KONV-2012-0017.

\appendix
\section{Widom particle insertion method to compute the components of the excess chemical potential}
\label{sec:wid}

The excess chemical potential profile can be computed using Widom's particle insertion method \cite{widom63,widom78}.
We divide the simulation cell into small volume elements as described in reference \cite{boda-jcp-134-055102-2011} and insert ``ghost'' particles into uniformly generated positions in these volume elements.
We compute the interaction energy $U(\mathbf{r})$ of the ``ghost'' ion inserted at position $\mathbf{r}$ with the whole system and use it in the operation
\begin{equation}
 \mathcal{W}\left[ U(\mathbf{r})\right]  = -kT \ln \left\langle \re^{-U(\mathbf{r})/kT}\right\rangle  ,
\label{eq:w-operation}
\end{equation}
where the brackets denote GC ensemble average.
If the interaction energy $U(\mathbf{r})$ contains all the terms (no matter how it is
divided into terms), operator $\mathcal{W}$ provides the full excess chemical potential
\begin{equation}
\mu_{i}^{\mathrm{EX}}(\mathbf{r}) = \mathcal{W}\left[ U_{i}^{\mathrm{HS}}(\mathbf{r})+ U_{i}^{\mathrm{II}}(\mathbf{r})+ U_{i}^{\mathrm{ID}}(\mathbf{r}) + U_{i}^{\mathrm{SELF}}(\mathbf{r}) \right].
\end{equation}
A diverging term $U_{i}^{\mathrm{WALL}}(\mathbf{r})$ corresponding to overlap with protein and membrane walls is omitted in this equation, because we evaluate the excess chemical potential only at the allowed positions.

The II term of the energy is obtained as
$U_{i}^{\mathrm{II}}(\mathbf{r})=\sum_{j\neq i} z_{i}z_{j}e^{2}\psi_{ij}^{\mathrm{II}}(\mathbf{r},\mathbf{r}_{j})$, where
\begin{equation}
\psi_{ij}^{\mathrm{II}}(\mathbf{r}_{i},\mathbf{r}_{j})= \dfrac{1}{8\pi \epsilon_{0} \epsilon_{\mathrm{w}} |\mathbf{r}_{i}-\mathbf{r}_{j}|}
\label{eq:psi_II}
\end{equation}
describes the Coulomb interaction between two unit charges at positions $\mathbf{r}_{i}$ and $\mathbf{r}_{j}$.
The ID term is obtained as
$U_{i}^{\mathrm{ID}}(\mathbf{r})=\sum_{j\ne i} z_{i}z_{j}e\psi_{ij}^{\mathrm{ID}}(\mathbf{r},\mathbf{r}_{j})$, where
\begin{equation}
 \psi_{ij}^{\mathrm{ID}} (\mathbf{r}_{i}, \mathbf{r}_{j} ) = \dfrac{1}{8\pi \epsilon_{0}} \left[
\int_{\mathcal{B}} \dfrac{h_{j}(\mathbf{r}_{j},\mathbf{s})}{|\mathbf{r}_{i}-\mathbf{s}|} \rd\mathbf{s} +
\int_{\mathcal{B}} \dfrac{h_{i}(\mathbf{r}_{i},\mathbf{s})}{|\mathbf{r}_{j}-\mathbf{s}|} \rd\mathbf{s} \right]
\label{eq:psi_ID}
\end{equation}
describes the interaction of a unit charge at $\mathbf{r}_{i}$ with the polarization charge, $h_{j}(\mathbf{r}_{j},\mathbf{s})$, induced by another unit charge at $\mathbf{r}_{j}$ (or vice versa).
Vector $\mathbf{s}$ runs over the dielectric boundary $\mathcal{B}$.
The polarization charge is determined using our Induced Charge Computation method \cite{boda-pre-69-046702-2004,boda-jcp-125-034901-2006}.

We define the terms in the excess chemical potential that correspond to different interactions as suggested by Gillespie \cite{gillespie-bj-2008}.
The definition of these terms is not unique.
In our previous work \cite{boda-jcp-134-055102-2011}, we suggested a possible and physically well-based procedure.
The HS term in the excess chemical potential is computed by inserting uncharged hard spheres into the system with the same size as the corresponding ion, but without the charge:
\begin{equation}
\mu_{i}^{\mathrm{HS}} (\mathbf{r})=   \mathcal{W}\left[ U_{i}^{\mathrm{HS}} (\mathbf{r})\right] .
\end{equation}
The $\mathrm{II}+\mathrm{ID}+\mathrm{SELF}$ part is the difference $\mathrm{EX}-\mathrm{HS}$.
If we insert charged hard spheres into the system, but ignore their interactions with the polarization charges, we can compute an excess chemical potential term describing the ion-ion interactions including the HS interactions: $\mathcal{W}\left[ U_{i}^{\mathrm{HS}} (\mathbf{r}) +U_{i}^{\mathrm{II}} (\mathbf{r}) \right]$.
The II term (that corresponds solely to the interaction with the ionic charges) is obtained by subtracting the HS term:
\begin{equation}
 \mu_{i}^{\mathrm{II}} (\mathbf{r}) =  \mathcal{W}\left[ U_{i}^{\mathrm{HS}} (\mathbf{r}) +U_{i}^{\mathrm{II}} (\mathbf{r}) \right] -  \mathcal{W}\left[ U_{i}^{\mathrm{HS}} (\mathbf{r})\right].
\end{equation}
The ID term (that corresponds to the interactions with polarization charges induced by other ions) is what remains:
\begin{equation}
\mu_{i}^{\mathrm{ID}}(\mathbf{r}) = \mu_{i}^{\mathrm{EX}} (\mathbf{r}) -  \mu_{i}^{\mathrm{HS}} (\mathbf{r}) - \mu_{i}^{\mathrm{II}} (\mathbf{r}) -\mu^{\mathrm{SELF}}_{i}(\mathbf{r}).
\end{equation}
The SELF term is a one-particle term that corresponds to the $i=j$ term of the ID energy in equation (\ref{eq:psi_ID}).
The MF term is simply the interaction with the mean electric field computed by sampling with a unit point charge as described in reference \cite{boda-jcp-134-055102-2011}.
The SC term, again, is what remains: $\mathrm{SC} = \mathrm{EX} - \mathrm{HS} - \mathrm{MF} - \mathrm{SELF}$.

\ukrainianpart

\title{Енергетика іонної конкуренції у селективному фільтрі DEKA нейронних натрієвих каналів}
 \author[D. Boda \textsl{et al.}]{Д. Бода\refaddr{label4}, Г. Ліф\refaddr{label1}, Дж. Фонсека\refaddr{label2}, Б. Айзенберг\refaddr{label3}}
 \addresses{
 \addr{label4} Факультет фізичної хімії, Університет Паннонії, Веспрем, H-8201, Угорщина
 \addr{label1} Відділення математики і комп'ютерних наук, Аргоннська національна лабораторія, \\ м. Аргонн, IL 60439, США
 \addr{label2} Мережа обчислювальної нанотехнології, Університет Пердью, м. Вест-Лафайетт, штат Індіана, США
 \addr{label3} Факультет молекулярної біофізики і фізіології, Медичний центр університету Раша,  \\ вул. Вест Гаррісон, 1750, Чікаго, США
 }

 \makeukrtitle

\begin{abstract}

 Проведено дослідження енергетики іонної селективності в нейронних каналах натрію. Використано просту модель,
 сконструйовану спеціально для селективного фільтру. Селективний фільтр канального типу містить залишки аспарату (D),
 глютамату (E), лізину K) та аланіну (A) (область DEKA). Використано моделювання методом Монте Карло у великому
 канонічному ансамблі для обчислення селективності рівноважного зв'язування у селективному фільтрі і для отримання
 різних членів надлишкового хімічного потенціалу в результаті процедури вставляння частинок на основі методу Відома.
 Показано, що іони K$^{+}$ у суперництві з Na$^{+}$ ефективно вилучаються з селективного фільтра за рахунок ентропійного
 виключен\-ня твердих сфер. Діелектрична проникність протеїну не має жодного впливу на дану селективність. З іншого боку,
 іони Ca$^{2+}$ вилучаються з фільтра за рахунок
 вільного енергетичного ``пенальті'', що підсилюється низькою діелектричною проникністю протеїну.

\keywords Монте Карло, примітивна модель електролітів, іонний канал, селективність
 \end{abstract}

\end{document}